\documentclass[twocolumn,showpacs,preprintnumbers,amsmath,amssymb]{revtex4-1}

\usepackage{amssymb}
\usepackage{amsmath}
\usepackage{graphicx}
\usepackage{color}

\newcommand{\be}{\begin{equation}}
\newcommand{\ee}{\end{equation}}
\newcommand{\bea}{\begin{eqnarray}}
\newcommand{\eea}{\end{eqnarray}}

\begin{document}

\preprint{DCPT-15/03}


\title{Gravity and the stability of the Higgs vacuum}


\author{Philipp Burda$^1$}
\author{Ruth Gregory$^{1,2}$}
\author{Ian G.\ Moss$^3$}
\affiliation{$^1$Centre for Particle Theory, Durham University,
South Road, Durham, DH1 3LE, UK\\
$^2$Perimeter Institute, 31 Caroline St, Waterloo, Ontario N2L 2Y5,
Canada\\
$^3$School of Mathematics and Statistics, Newcastle University,
Newcastle Upon Tyne, NE1 7RU, U.K.}


\date{\today}

\begin{abstract}
We discuss the effect of gravitational interactions on the lifetime
of the Higgs vacuum where generic quantum gravity corrections 
are taken into account. Using a ``thin-wall'' approximation, we 
provide a proof of principle that small black holes can act as 
seeds for vacuum decay, spontaneously nucleating a new Higgs 
phase centered on the black hole with a lifetime measured in millions 
of Planck times rather than billions of years.
The corresponding parameter space constraints are extremely 
stringent however, therefore we also present numerical evidence 
suggesting that with thick walls, the parameter space may open up.
Implications for collider black holes are discussed.
\end{abstract}


\maketitle


With the discovery of the Higgs boson at the LHC
\cite{ATLAS:2012ae,Chatrchyan:2012tx}, and the measurement of
its mass, it seems we live in interesting times: the running of
the coupling of the Higgs quite possibly means that our vacuum is
only metastable, and the true Higgs vacuum in fact lies at large
expectation values of the Higgs and negative vacuum energy.
Although this metastability at first might seem alarming, in order
for the vacuum to decay, it must tunnel through a sizeable energy
barrier, and the probability for this typically has an exponential factor,
\begin{equation}
\Gamma \sim A e^{-B/\hbar}
\label{decayaction}
\end{equation}
where $B$ is the action of a solution to the Euclidean field
equations, ``the bounce'', which interpolates between the metastable
(false) and true vacua; the prefactor $A$ is determined from 
fluctuations around the bounce. Since the action $B$ is usually
large (we will set $\hbar$ to 1 for the rest of this discussion) 
the probability of vacuum decay is very low. For the decay of
a false vacuum, the process was understood and the probability
computed in a series of papers by Coleman, Callan and de Luccia 
\cite{Coleman:1977py,Callan:1977pt,Coleman:1980aw}.
This `gold standard' calculation is now used ubiquitously to estimate decay
rates and the half life of a false vacuum state in field theory,
and for the Higgs vacuum predicts a lifetime well in excess of 
the age of the universe.

The Coleman et al.\ picture of vacuum decay is however very idealized,
in that an exactly homogeneous and isotropic false vacuum decays 
into a very nearly as symmetric configuration: a completely spherical 
bubble of true vacuum which expands outwards with uniform acceleration.
In everyday physics however, first order phase transitions are far from 
clean, and often proceed not via some perfect nucleation process
but rather by impurities acting as sites for the condensation of a new
phase. Recently in \cite{Gregory:2013hja} we investigated the impact 
of gravitational impurities, in the guise of black holes, on the usual
Coleman de Luccia picture (see also \cite{Hiscock:1987hn,Berezin:1990qs} 
for early investigations of this issue) and found a significant enhancement
of vacuum decay in a wide range of cases -- the intuition of an impurity 
seeding nucleation of
a true vacuum bubble was entirely borne out by the analysis.

To recall Coleman's original intuition: the nucleation of a bubble costs
energy because a wall with energy and tension is formed as a barrier
between the false and true vacua, but to counter that, energy is gained 
by the volume of space inside the bubble having lower energy by virtue
of having transitioned to the true vacuum. A bubble of just the right size 
then optimises
this energy pay-off and once formed, grows. The picture with a
gravitational inhomogeneity is similar; the bubble forms around
the (euclidean) black hole but because of the distortion of space
the payoff between volume inside a bubble and its surface area
is changed, and bubbles form at a smaller radius and hence the 
`cost' of forming them is lower: the instanton has a smaller action 
and the decay process can be significantly enhanced.
Alternatively, in terms of the original energy argument of Coleman
and de Luccia, the addition of a seed black hole which is eliminated
or reduced by the bubble can change the energy balance dramatically.

Can this process affect the lifetime of the Higgs vacuum?
We will show that it can, although only if small black holes
nucleate the decay. Such black holes could result from gradual 
evaporation of primordial black holes formed in the early 
universe \cite{Carr:1974nx}; alternatively, if there are ``large''
extra dimensions \cite{ArkaniHamed:1998rs,Randall:1999ee}
responsible for producing a hierarchically 
large Planck scale in our universe, then small black holes can 
be produced at the LHC \cite{Dimopoulos:2001hw}. 
Depending on the tension of the
bubble wall, which is directly related to parameters in the
Higgs potential, the enhancement of vacuum decay can
be large.

To briefly review the Higgs potential, note that
at large values of the Higgs field, we can pick any
component $\phi$ and approximate the potential
using an effective coupling constant $\lambda_{\rm eff}$,
\begin{equation}
V(\phi)=\frac14\lambda_{\rm eff}(\phi)\phi^4.
\end{equation}
The effective coupling is obtained by combining the running
of $\lambda$ under the renormalisation group
with the low-energy particle physics parameters. Two-loop
calculations of the running coupling \cite{Degrassi:2012ry},
including contributions from all of the standard model fields, 
yield a high-energy approximation 
\begin{equation}
\lambda_{\rm eff}\approx\lambda_*+b\left(\ln{\phi\over\phi_*}\right)^2,
\end{equation}
where the fit to the results of \cite{Degrassi:2012ry} give
parameter ranges $-0.01\lesssim\lambda_*\lesssim0$,
$0.1M_p\lesssim \phi_*\lesssim M_p$ and $b\sim 10^{-4}$. 
These parameters ranges are mostly due to experimental 
uncertainties in the Higgs and top quark mass, however 
with the currently measured values it seems
that  $\lambda_{\rm eff}$ near the Planck scale is
small with a preference towards negative values.

\begin{figure}
\includegraphics[scale=0.43]{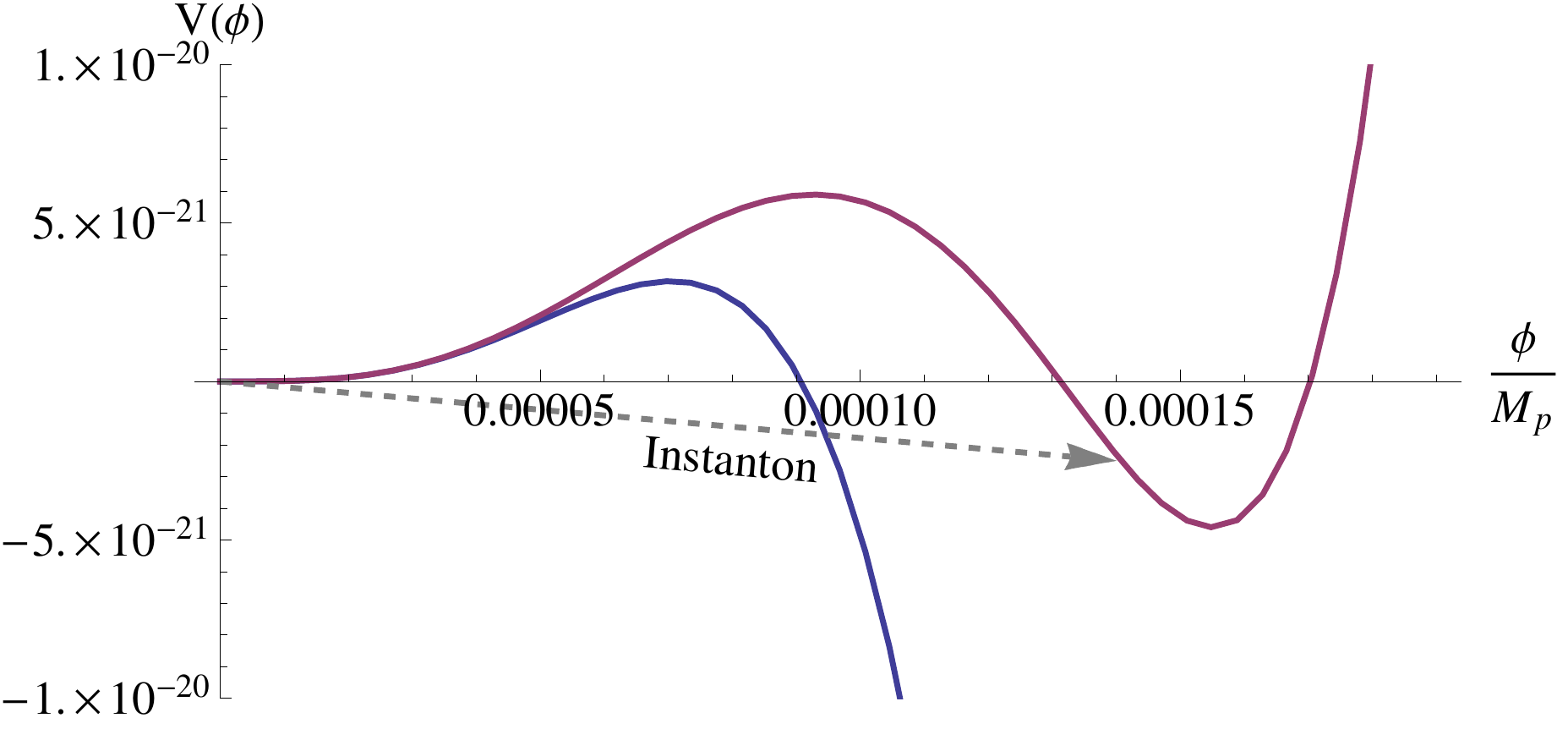}
\caption{An illustration of the impact of the quantum gravity 
correction (in red), here shown for $\lambda_*=-0.01, \phi_*=2M_p$, and
$\lambda_6=63K$. The path of the bounce is sketched.
\label{fig:higgspot}
}
\end{figure}

Of course this discussion assumes no impact from new physics 
between the TeV scale and the Planck scale. At the very least, 
quantum gravity effects will have to be taken into account. On
dimensional grounds, we might expect some modifications
to the potential of the following form \cite{Moss:2014nya}:
\begin{equation}
V(\phi)=\lambda_{\rm eff}(\phi)\frac{\phi^4}4
+(\delta\lambda)_{\rm bsm}\frac{\phi^4}4
+\frac{\lambda_6}{6}{\phi^6\over M_p^2}
+\frac{\lambda_8}{8}{\phi^8\over M_p^4}+...
\label{hehiggs}
\end{equation}
where $(\delta\lambda)_{\rm bsm}$ includes possible
corrections to the running coupling from physics
beyond the standard model, and the polynomial terms 
arise with new physics identified with the Planck scale. 
If the coefficients $\lambda_6$ etc.\  are similar in
magnitude, then the small size of $\lambda_{\rm eff}$
at the Planck scale means that the interesting physics
occurs where the potential is
determined predominantly by $\lambda_{\rm eff}$
and $\lambda_6$. In figure \ref{fig:higgspot} we illustrate
the effect of these corrections on the standard
model potential with $\lambda_*=-0.01$.

Quantum tunnelling in a corrected potential such as this
has been looked at by Branchina et al.\
\cite{Branchina:2013jra}, who
take $\lambda_*\sim-0.1$, where
the potential barrier occurs at $\phi\ll M_p$, and take a
negative $\lambda_6=-2$. They argue the existence of
a greatly enhanced tunnelling rate, 
however, their discussion entirely neglected gravitational
back-reaction of the instanton on the geometry. 

In order to explore the impact of a gravitational impurity
we will extend the method of Coleman and de 
Luccia (CDL) \cite{Coleman:1980aw}, to include a black 
hole. We find that within the CDL ``thin-wall'' description, 
the tunneling amplitude can be significantly enhanced 
by a small black hole, albeit within a small region of
parameter space. This provides a `proof of principle'
and motivates a numerical analysis of Higgs instantons,
which confirm the presence of strongly enhanced decay 
in the presence of black holes.

The equations of motion for a
wall bounding two different regions of spacetime with
different cosmological constants and black hole masses can
be expressed in the form $\dot R^2+V(R)=0$, where
$R$ is the bubble radius (as a function of Euclidean time), and
$V(R)$ is an effective potential involving the wall tension.
For the decay of the Higgs vacuum, we assume the standard
model has $\Lambda_+=0$, and write the true vacuum
cosmological constant as $\Lambda_- = -3/\ell^2$, then the
potential $V$ depends on $\ell$, the black hole masses $M_\pm$
and the surface tension of the bubble wall. (See \cite{Sasaki:2014spa} 
for explicit forms of this potential.)

To recapitulate the results of \cite{Gregory:2013hja},
the action of a general instanton with a black hole
was found to be 
\begin{eqnarray}
B=&& {{\cal A}_h^+\over 4G}-{{\cal A}_h^-\over 4G}
\label{mplusminusaction}\\
&&+{1\over 4G}\oint d\lambda \left\{\left ( 2R - 6GM_+\right)\dot{\tau}_+
- \left (2R-6GM_-\right) \dot{\tau}_-\right\}\nonumber
\end{eqnarray}
where $R(\tau_\pm)$ is the solution for the bubble wall,
and ${\cal A}_h^\pm$ are the black hole horizon areas corresponding
to $M_\pm$. This result includes a careful treatment of the 
conical deficits which can arise in the Euclidean section
when the periodicity of the bubble solution is not the same
as that of the black hole, and although the specifics of computing
actions in vacuum and AdS vary from that of dS, the essence of
the calculation remains the same as the presentation in 
\cite{Gregory:2013hja}, and the result, \eqref{mplusminusaction},
identical in form.

This bounce action feeds directly into the exponent in \eqref{decayaction},
and following Callan and Coleman \cite{Callan:1977pt}, we estimate
the prefactor by taking a factor of $(B/2\pi)^{1/2}$ for the single
time-translational zero mode of the instanton but use the light crossing
time of the black hole, $(GM_+)^{-1}$, as a rough estimate of
the remaining determinant of fluctuations giving
\begin{equation}
\Gamma_D\approx \left({B\over 2\pi GM_+}\right)^{1/2}e^{-B}.
\label{decayamplitude}
\end{equation}
Typically, the CDL action is of order ${\cal O}(10^{3-6})$ for the Higgs
potentials, leading to a huge exponential suppression of the decay
rate, and to the conclusion that gravitational tunneling is irrelevant.

However, the effect of a black hole, \eqref{mplusminusaction},
on the tunneling action can be very significant for low tension 
bubble walls and small mass black holes. As the seed black hole 
mass $M_+$ is switched on, the instanton action drops rapidly,
and the bubble initially nucleates by removing the black hole.  
However, as the seed black hole mass continues to increase, 
a critical mass $M_C$ is reached at which the potential $V(R)$ 
has a single point at which $V=V'=0$, and there exists a static
bubble wall solution. In this case, an unstable static bubble
nucleates which will either recollapse or expand
with roughly equal probability. As the seed black hole mass 
increases further, the nucleated bubble now has
a black hole remnant in the bubble interior, with the
action now rising with increasing seed mass.
The quantitative values of this critical mass, and the maximal
suppression of the bounce action at $M_C$ depend on the wall 
tension parameter $\sigma$, and the true vacuum energy,
however, unless the combination $\sigma\ell$ is Planck scale,
this suppression is several orders of magnitude at $M_C$, thus 
changing the exponential factor in \eqref{decayamplitude}
from an irrelevant $10^6$ to a potentially extremely 
relevant $10^{0-2}$.

Whether or not this enhancement is relevant depends on
its magnitude relative to other physical decay processes,
specifically, black hole evaporation.
The key indicator is therefore the branching ratio of the static 
tunneling decay rate to the Hawking evaporation rate, 
$\Gamma_H\approx 3.6\times 10^{-4}(G^2 M_+^3)^{-1}$
\cite{Page:1976df}:
\begin{equation}
\Gamma_D/\Gamma_H\approx
44(M_+^2/ M_p^2)B^{1/2}e^{-B}.
\label{bratio}
\end{equation}
For our thin wall instantons, there is indeed a range of
$M_+$ (small, though still above the Planck mass),
for which we have very strong
enhancement of bubble tunneling.

The main wrinkle in this argument is that the condition for 
the thin wall approximation requires that the energy at
the potential minimum is smaller than the potential
barrier height, and scanning through parameter space 
we find that requiring a thin wall is very constraining: 
the range of $\lambda_6$ for which this occurs is very small,
and occurs for large values of the parameter $\lambda_6 
\gtrsim 10^3-10^5$, depending on $\lambda_*$. On the other
hand, computing the branching ratio, \eqref{bratio}, for
these models shows that tunneling does indeed dominate.
Thus, while our pseudo-analytic discussion is limited in the sense 
of parameter space, it has provided a {\it proof of principle} 
that black holes could potentially seed vacuum decay. 

In order to decide whether this effect is restricted to a niche 
of parameter space, or is potentially relevant, a full exploration 
of instantons outside of the thin wall approximation is necessary. 
Motivated by our thin wall results, in which the enhanced tunneling 
takes place with the static instanton (as $M_+>M_C$, which is 
typically less than the Planck mass), we have made a preliminary 
numerical investigation of static instantons, taking 
$\lambda_*=-0.01$, and $b=10^{-4}$ as representative
values for the Higgs potential.

\begin{figure}
\includegraphics[scale=0.4]{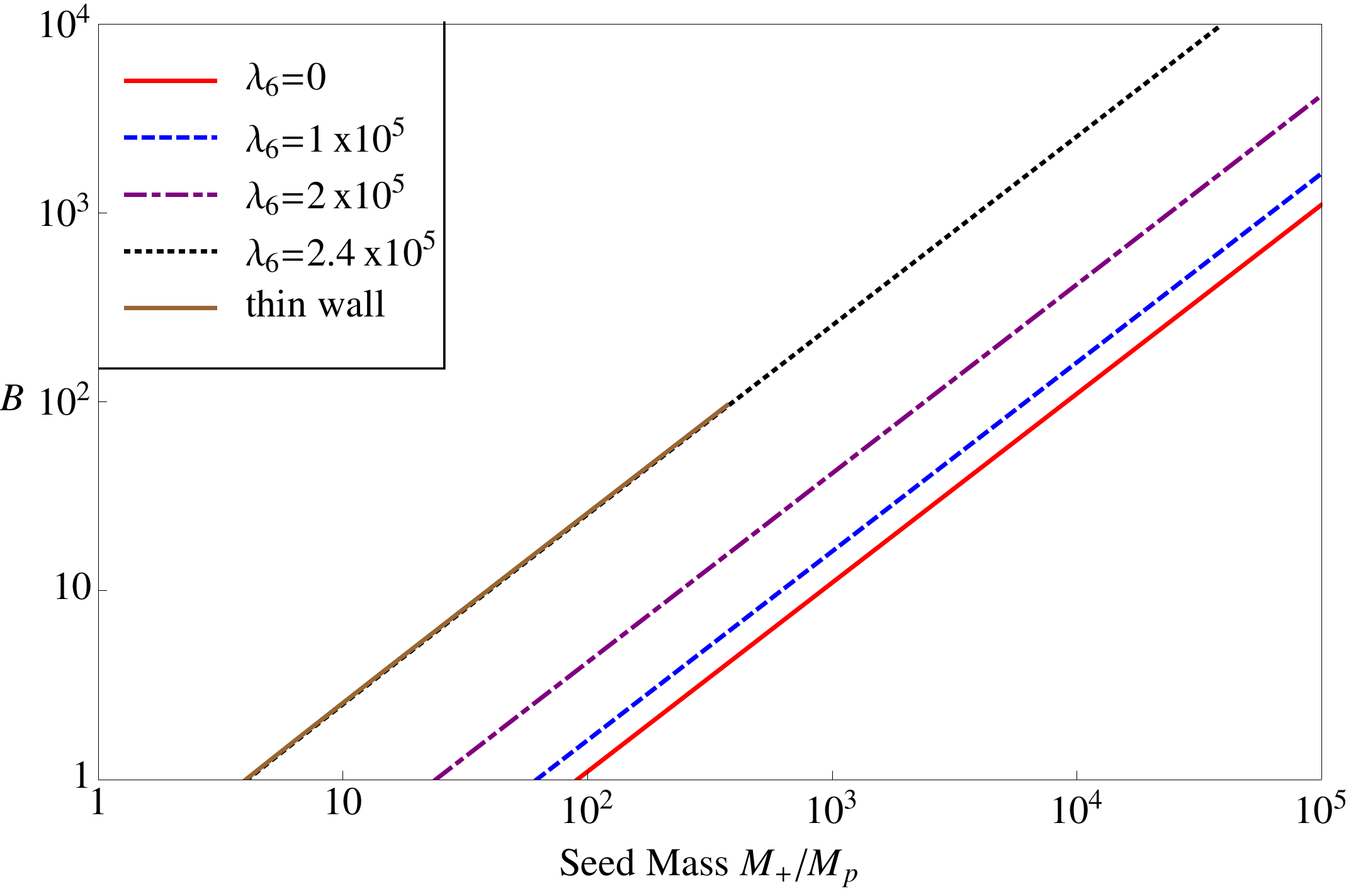}
\caption{
The action for a bounce solution. Each plot corresponds to
a different value of $\lambda_6$ in the Higgs potential
\eqref{hehiggs}, with 
$\lambda_*=-0.01$ and $b=1.0\times 10^{-4}$. 
The largest value of $\lambda_6$ is within the range of
the thin wall approximation, and the thin wall result is
shown for comparison.
\label{fig:action}
}
\end{figure}

Static bounce solutions to the Einstein-scalar equations with 
rotational symmetry on a black hole AdS background can be 
found using a spherically symmetric metric ansatz
\begin{equation}
ds^2=f(r)e^{2\delta(r)}d\tau^2+{dr^2\over f(r)}
+r^2(d\theta^2+\sin^2\theta d\varphi^2),
\end{equation}
where
\begin{equation}
f=1-{2G\mu(r)\over r}.
\end{equation}
The solutions are obtained using a shooting technique, 
varying the value of the scalar field at the black hole horizon 
and aiming for $\phi\to0$ as $r\to 0$. In Ref. \cite{Gregory:2013hja}, 
it was shown that the action is given by the area terms in 
\eqref{mplusminusaction}, as in the thin wall case. The resulting 
values of the action for a selection of Higgs models is shown in figure 
\ref{fig:action}. Note that the semi-classical bubble nucleation argument 
only applies when the action $B>1$. 

\begin{figure}
\includegraphics[scale=0.35]{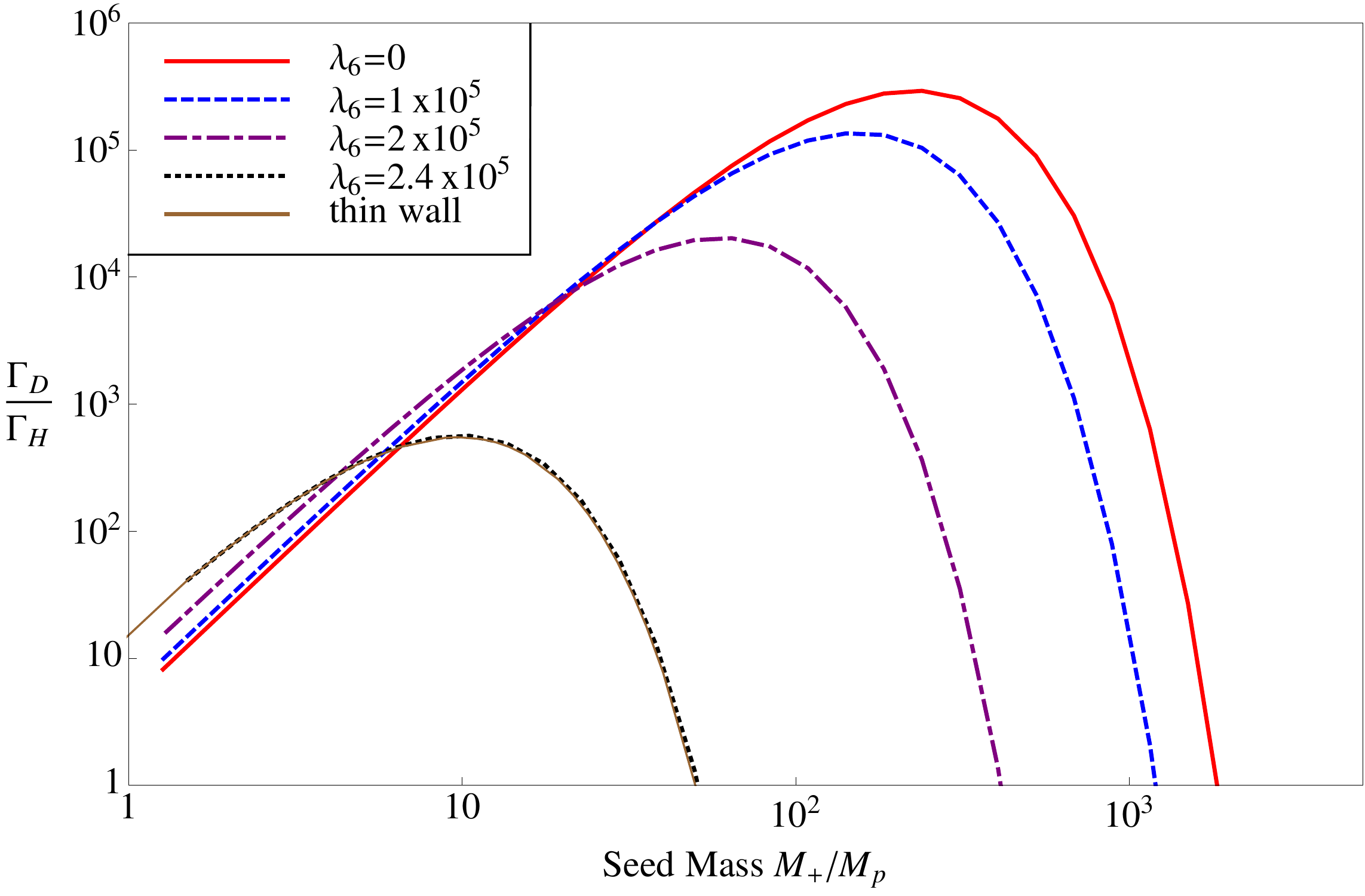}
\caption{The branching ratio of the false vacuum nucleation rate 
to the Hawking evaporation rate as a function 
of the seed mass for a selection of Higgs models. Each
plot corresponds a different value of $\lambda_6$ in \eqref{hehiggs},
with $\lambda_*=-0.01$.
\label{fig:branchratio}
}
\end{figure}

Computing the branching ratio now with these ``thick wall''
solutions gives figure \ref{fig:branchratio}. Although 
black holes produced in the early universe start out with 
relatively high masses, their temperature is nonetheless above 
that of the microwave background, and they evaporate down
into the range plotted in figure \ref{fig:branchratio}. 
At this point, the mass hits a range in which vacuum decay is
more probable, i.e.\ the tunneling half life becomes smaller
than the (instantaneous) Hawking lifetime of the black hole. 
Note that this range is well above the 
Planck mass, where we have some
confidence in the validity of the vacuum decay calculation.
Given that this evaporation timescale is $\sim 10^{-28}$s for
a $10^5M_p$ mass black hole, it is clear that once a primordial
black hole nears the end of its life cycle, it {\it will} seed
vacuum decay in these models. 
Hence with these Higgs potentials,
the presence of any primordial black holes will eventually
trigger a catastrophic phase transition from our standard model
vacuum thus ruling out potentials with parameters in these ranges.

Since our results show that it is precisely for small black holes
that the risk of seeded tunneling is greatest, a natural question
is what happens with collider black holes. These can be
produced if the fundamental (higher dimensional)
Planck scale is near the TeV scale \cite{Dimopoulos:2001hw}.
These black holes have features inherited from their
higher dimensional nature, and while there are no known
exact solutions, evaporation rates have 
been computed assuming a higher dimensional 
Myers-Perry solution \cite{Myers:1986un}, with 
emission cross sections appropriate to a braneworld
scenario \cite{Kanti:2014vsa}.

Black hole seeded tunneling is now a more involved process,
as it should involve a bubble forming around the 
higher dimensional black hole triggered by the Higgs
field transitioning on the brane, the bubble then expanding 
out to fill the extra dimensions before finally becoming 
effectively four-dimensional and seeding true decay of our universe. 
While this process is beyond the reach of the analytic
approximations we have used here, we can estimate the effect
by modelling the instanton with a higher dimensional counterpart
of the solutions described above. In this case, the form of the
potential $V(R)$ for the bubble motion is modified, but of a
remarkably similar form, essentially replacing $R\to R^{n+1}$,
where $n$ is the number of extra dimensions. Assuming the
static bubble we can then calculate the horizon radius and area: the 
action will be the difference in seed and remnant black hole
horizon areas. It turns out this calculation is relatively insensitive
to the number of extra dimensions (the horizon areas ${\cal A}\propto
M^{(n+2)/(n+1)}$) whereas the evaporation rate of black holes is 
enhanced, in part because of the increased Hawking temperature,
$T\propto M^{-1/(n+1)}$, and in part because of grey-body factors.
The branching ratio tends to be suppressed with extra dimensions, 
making collider black holes less risky for vacuum decay, however
black holes produced by particle collisions could still cause 
vacuum decay in certain regions of parameter space. 
Fortunately, we have some reassurance about the safety
of the LHC from the fact that cosmic ray collisions have occurred 
at energies higher than those reached at the collider \cite{Hut1983}! 

To sum up: We have shown that the Coleman de Luccia result
for the lifetime of our universe in Higgs potentials with metastability
seems crucially dependent on the absence of inhomogeneities - the
presence of primordial black holes can dramatically reduce the
barrier to vacuum decay, and seed nucleation to a universe
with a very different ``standard model''.
Such a conclusion of course depends on the existence of said
small black holes -- by no means a certainty -- and a
detailed numerical study of parameter space. However,
these results are suggestive that the issue of metastability of 
our universe may not be as
simple as initially was thought.

\begin{acknowledgments}
We would like to thank Ben Withers for collaboration in the early stages
of this project, and Yiota Kanti for comments on the manuscript.
PB is supported by an EPSRC International Doctoral Scholarship,
RG and IGM are supported in part by STFC (Consolidated Grant ST/L000407/1).
RG is also supported by the Wolfson Foundation and Royal Society, and
Perimeter Institute for Theoretical Physics.
Research at Perimeter Institute is supported by the Government of
Canada through Industry Canada and by the Province of Ontario through the
Ministry of Research and Innovation.

\end{acknowledgments}


\begin{thebibliography}{20}%
\makeatletter
\providecommand \@ifxundefined [1]{%
 \@ifx{#1\undefined}
}%
\providecommand \@ifnum [1]{%
 \ifnum #1\expandafter \@firstoftwo
 \else \expandafter \@secondoftwo
 \fi
}%
\providecommand \@ifx [1]{%
 \ifx #1\expandafter \@firstoftwo
 \else \expandafter \@secondoftwo
 \fi
}%
\providecommand \natexlab [1]{#1}%
\providecommand \enquote  [1]{``#1''}%
\providecommand \bibnamefont  [1]{#1}%
\providecommand \bibfnamefont [1]{#1}%
\providecommand \citenamefont [1]{#1}%
\providecommand \href@noop [0]{\@secondoftwo}%
\providecommand \href [0]{\begingroup \@sanitize@url \@href}%
\providecommand \@href[1]{\@@startlink{#1}\@@href}%
\providecommand \@@href[1]{\endgroup#1\@@endlink}%
\providecommand \@sanitize@url [0]{\catcode `\\12\catcode `\$12\catcode
  `\&12\catcode `\#12\catcode `\^12\catcode `\_12\catcode `\%12\relax}%
\providecommand \@@startlink[1]{}%
\providecommand \@@endlink[0]{}%
\providecommand \url  [0]{\begingroup\@sanitize@url \@url }%
\providecommand \@url [1]{\endgroup\@href {#1}{\urlprefix }}%
\providecommand \urlprefix  [0]{URL }%
\providecommand \Eprint [0]{\href }%
\providecommand \doibase [0]{http://dx.doi.org/}%
\providecommand \selectlanguage [0]{\@gobble}%
\providecommand \bibinfo  [0]{\@secondoftwo}%
\providecommand \bibfield  [0]{\@secondoftwo}%
\providecommand \translation [1]{[#1]}%
\providecommand \BibitemOpen [0]{}%
\providecommand \bibitemStop [0]{}%
\providecommand \bibitemNoStop [0]{.\EOS\space}%
\providecommand \EOS [0]{\spacefactor3000\relax}%
\providecommand \BibitemShut  [1]{\csname bibitem#1\endcsname}%
\let\auto@bib@innerbib\@empty
\bibitem [{\citenamefont {Aad}\ \emph {et~al.}(2012)\citenamefont {Aad} \emph
  {et~al.}}]{ATLAS:2012ae}%
  \BibitemOpen
  \bibfield  {author} {\bibinfo {author} {\bibfnamefont {G.}~\bibnamefont
  {Aad}} \emph {et~al.} (\bibinfo {collaboration} {ATLAS Collaboration}),\
  }\href {\doibase 10.1016/j.physletb.2012.02.044} {\bibfield  {journal}
  {\bibinfo  {journal} {Phys.Lett.}\ }\textbf {\bibinfo {volume} {B710}},\
  \bibinfo {pages} {49} (\bibinfo {year} {2012})},\ \Eprint
  {http://arxiv.org/abs/1202.1408} {arXiv:1202.1408 [hep-ex]} \BibitemShut
  {NoStop}%
\bibitem [{\citenamefont {Chatrchyan}\ \emph {et~al.}(2012)\citenamefont
  {Chatrchyan} \emph {et~al.}}]{Chatrchyan:2012tx}%
  \BibitemOpen
  \bibfield  {author} {\bibinfo {author} {\bibfnamefont {S.}~\bibnamefont
  {Chatrchyan}} \emph {et~al.} (\bibinfo {collaboration} {CMS Collaboration}),\
  }\href {\doibase 10.1016/j.physletb.2012.02.064} {\bibfield  {journal}
  {\bibinfo  {journal} {Phys.Lett.}\ }\textbf {\bibinfo {volume} {B710}},\
  \bibinfo {pages} {26} (\bibinfo {year} {2012})},\ \Eprint
  {http://arxiv.org/abs/1202.1488} {arXiv:1202.1488 [hep-ex]} \BibitemShut
  {NoStop}%
\bibitem [{\citenamefont {Coleman}(1977)}]{Coleman:1977py}%
  \BibitemOpen
  \bibfield  {author} {\bibinfo {author} {\bibfnamefont {S.~R.}\ \bibnamefont
  {Coleman}},\ }\href {\doibase 10.1103/PhysRevD.15.2929,
  10.1103/PhysRevD.16.1248} {\bibfield  {journal} {\bibinfo  {journal}
  {Phys.Rev.}\ }\textbf {\bibinfo {volume} {D15}},\ \bibinfo {pages} {2929}
  (\bibinfo {year} {1977})}\BibitemShut {NoStop}%
\bibitem [{\citenamefont {Callan}\ and\ \citenamefont
  {Coleman}(1977)}]{Callan:1977pt}%
  \BibitemOpen
  \bibfield  {author} {\bibinfo {author} {\bibfnamefont {J.}~\bibnamefont
  {Callan}, \bibfnamefont {Curtis~G.}}\ and\ \bibinfo {author} {\bibfnamefont
  {S.~R.}\ \bibnamefont {Coleman}},\ }\href {\doibase 10.1103/PhysRevD.16.1762}
  {\bibfield  {journal} {\bibinfo  {journal} {Phys.Rev.}\ }\textbf {\bibinfo
  {volume} {D16}},\ \bibinfo {pages} {1762} (\bibinfo {year}
  {1977})}\BibitemShut {NoStop}%
\bibitem [{\citenamefont {Coleman}\ and\ \citenamefont
  {De~Luccia}(1980)}]{Coleman:1980aw}%
  \BibitemOpen
  \bibfield  {author} {\bibinfo {author} {\bibfnamefont {S.~R.}\ \bibnamefont
  {Coleman}}\ and\ \bibinfo {author} {\bibfnamefont {F.}~\bibnamefont
  {De~Luccia}},\ }\href {\doibase 10.1103/PhysRevD.21.3305} {\bibfield
  {journal} {\bibinfo  {journal} {Phys.Rev.}\ }\textbf {\bibinfo {volume}
  {D21}},\ \bibinfo {pages} {3305} (\bibinfo {year} {1980})}\BibitemShut
  {NoStop}%
\bibitem [{\citenamefont {Gregory}\ \emph {et~al.}(2014)\citenamefont
  {Gregory}, \citenamefont {Moss},\ and\ \citenamefont
  {Withers}}]{Gregory:2013hja}%
  \BibitemOpen
  \bibfield  {author} {\bibinfo {author} {\bibfnamefont {R.}~\bibnamefont
  {Gregory}}, \bibinfo {author} {\bibfnamefont {I.~G.}\ \bibnamefont {Moss}}, \
  and\ \bibinfo {author} {\bibfnamefont {B.}~\bibnamefont {Withers}},\ }\href
  {\doibase 10.1007/JHEP03(2014)081} {\bibfield  {journal} {\bibinfo  {journal}
  {JHEP}\ }\textbf {\bibinfo {volume} {1403}},\ \bibinfo {pages} {081}
  (\bibinfo {year} {2014})},\ \Eprint {http://arxiv.org/abs/1401.0017}
  {arXiv:1401.0017 [hep-th]} \BibitemShut {NoStop}%
\bibitem [{\citenamefont {Hiscock}(1987)}]{Hiscock:1987hn}%
  \BibitemOpen
  \bibfield  {author} {\bibinfo {author} {\bibfnamefont {W.~A.}\ \bibnamefont
  {Hiscock}},\ }\href {\doibase 10.1103/PhysRevD.35.1161} {\bibfield  {journal}
  {\bibinfo  {journal} {Phys. Rev.}\ }\textbf {\bibinfo {volume} {D35}},\
  \bibinfo {pages} {1161} (\bibinfo {year} {1987})}\BibitemShut {NoStop}%
\bibitem [{\citenamefont {Berezin}\ \emph {et~al.}(1991)\citenamefont
  {Berezin}, \citenamefont {Kuzmin},\ and\ \citenamefont
  {Tkachev}}]{Berezin:1990qs}%
  \BibitemOpen
  \bibfield  {author} {\bibinfo {author} {\bibfnamefont {V.~A.}\ \bibnamefont
  {Berezin}}, \bibinfo {author} {\bibfnamefont {V.~A.}\ \bibnamefont {Kuzmin}},
  \ and\ \bibinfo {author} {\bibfnamefont {I.~I.}\ \bibnamefont {Tkachev}},\
  }\href {\doibase 10.1103/PhysRevD.43.3112} {\bibfield  {journal} {\bibinfo
  {journal} {Phys. Rev.}\ }\textbf {\bibinfo {volume} {D43}},\ \bibinfo {pages}
  {3112} (\bibinfo {year} {1991})}\BibitemShut {NoStop}%
\bibitem [{\citenamefont {Carr}\ and\ \citenamefont
  {Hawking}(1974)}]{Carr:1974nx}%
  \BibitemOpen
  \bibfield  {author} {\bibinfo {author} {\bibfnamefont {B.~J.}\ \bibnamefont
  {Carr}}\ and\ \bibinfo {author} {\bibfnamefont {S.}~\bibnamefont {Hawking}},\
  }\href@noop {} {\bibfield  {journal} {\bibinfo  {journal}
  {Mon.Not.Roy.Astron.Soc.}\ }\textbf {\bibinfo {volume} {168}},\ \bibinfo
  {pages} {399} (\bibinfo {year} {1974})}\BibitemShut {NoStop}%
\bibitem [{\citenamefont {Arkani-Hamed}\ \emph {et~al.}(1998)\citenamefont
  {Arkani-Hamed}, \citenamefont {Dimopoulos},\ and\ \citenamefont
  {Dvali}}]{ArkaniHamed:1998rs}%
  \BibitemOpen
  \bibfield  {author} {\bibinfo {author} {\bibfnamefont {N.}~\bibnamefont
  {Arkani-Hamed}}, \bibinfo {author} {\bibfnamefont {S.}~\bibnamefont
  {Dimopoulos}}, \ and\ \bibinfo {author} {\bibfnamefont {G.}~\bibnamefont
  {Dvali}},\ }\href {\doibase 10.1016/S0370-2693(98)00466-3} {\bibfield
  {journal} {\bibinfo  {journal} {Phys.Lett.}\ }\textbf {\bibinfo {volume}
  {B429}},\ \bibinfo {pages} {263} (\bibinfo {year} {1998})},\ \Eprint
  {http://arxiv.org/abs/hep-ph/9803315} {arXiv:hep-ph/9803315 [hep-ph]}
  \BibitemShut {NoStop}%
\bibitem [{\citenamefont {Randall}\ and\ \citenamefont
  {Sundrum}(1999)}]{Randall:1999ee}%
  \BibitemOpen
  \bibfield  {author} {\bibinfo {author} {\bibfnamefont {L.}~\bibnamefont
  {Randall}}\ and\ \bibinfo {author} {\bibfnamefont {R.}~\bibnamefont
  {Sundrum}},\ }\href {\doibase 10.1103/PhysRevLett.83.3370} {\bibfield
  {journal} {\bibinfo  {journal} {Phys.Rev.Lett.}\ }\textbf {\bibinfo {volume}
  {83}},\ \bibinfo {pages} {3370} (\bibinfo {year} {1999})},\ \Eprint
  {http://arxiv.org/abs/hep-ph/9905221} {arXiv:hep-ph/9905221 [hep-ph]}
  \BibitemShut {NoStop}%
\bibitem [{\citenamefont {Dimopoulos}\ and\ \citenamefont
  {Landsberg}(2001)}]{Dimopoulos:2001hw}%
  \BibitemOpen
  \bibfield  {author} {\bibinfo {author} {\bibfnamefont {S.}~\bibnamefont
  {Dimopoulos}}\ and\ \bibinfo {author} {\bibfnamefont {G.~L.}\ \bibnamefont
  {Landsberg}},\ }\href {\doibase 10.1103/PhysRevLett.87.161602} {\bibfield
  {journal} {\bibinfo  {journal} {Phys.Rev.Lett.}\ }\textbf {\bibinfo {volume}
  {87}},\ \bibinfo {pages} {161602} (\bibinfo {year} {2001})},\ \Eprint
  {http://arxiv.org/abs/hep-ph/0106295} {arXiv:hep-ph/0106295 [hep-ph]}
  \BibitemShut {NoStop}%
\bibitem [{\citenamefont {Degrassi}\ \emph {et~al.}(2012)\citenamefont
  {Degrassi}, \citenamefont {Di~Vita}, \citenamefont {Elias-Miro},
  \citenamefont {Espinosa}, \citenamefont {Giudice} \emph
  {et~al.}}]{Degrassi:2012ry}%
  \BibitemOpen
  \bibfield  {author} {\bibinfo {author} {\bibfnamefont {G.}~\bibnamefont
  {Degrassi}}, \bibinfo {author} {\bibfnamefont {S.}~\bibnamefont {Di~Vita}},
  \bibinfo {author} {\bibfnamefont {J.}~\bibnamefont {Elias-Miro}}, \bibinfo
  {author} {\bibfnamefont {J.~R.}\ \bibnamefont {Espinosa}}, \bibinfo {author}
  {\bibfnamefont {G.~F.}\ \bibnamefont {Giudice}},  \emph {et~al.},\ }\href
  {\doibase 10.1007/JHEP08(2012)098} {\bibfield  {journal} {\bibinfo  {journal}
  {JHEP}\ }\textbf {\bibinfo {volume} {1208}},\ \bibinfo {pages} {098}
  (\bibinfo {year} {2012})},\ \Eprint {http://arxiv.org/abs/1205.6497}
  {arXiv:1205.6497 [hep-ph]} \BibitemShut {NoStop}%
\bibitem [{\citenamefont {Moss}(2014)}]{Moss:2014nya}%
  \BibitemOpen
  \bibfield  {author} {\bibinfo {author} {\bibfnamefont {I.~G.}\ \bibnamefont
  {Moss}},\ }\href@noop {} {\  (\bibinfo {year} {2014})},\ \Eprint
  {http://arxiv.org/abs/1409.2108} {arXiv:1409.2108 [hep-th]} \BibitemShut
  {NoStop}%
\bibitem [{\citenamefont {Branchina}\ and\ \citenamefont
  {Messina}(2013)}]{Branchina:2013jra}%
  \BibitemOpen
  \bibfield  {author} {\bibinfo {author} {\bibfnamefont {V.}~\bibnamefont
  {Branchina}}\ and\ \bibinfo {author} {\bibfnamefont {E.}~\bibnamefont
  {Messina}},\ }\href {\doibase 10.1103/PhysRevLett.111.241801} {\bibfield
  {journal} {\bibinfo  {journal} {Phys.Rev.Lett.}\ }\textbf {\bibinfo {volume}
  {111}},\ \bibinfo {pages} {241801} (\bibinfo {year} {2013})},\ \Eprint
  {http://arxiv.org/abs/1307.5193} {arXiv:1307.5193 [hep-ph]} \BibitemShut
  {NoStop}%
\bibitem [{\citenamefont {Sasaki}\ and\ \citenamefont
  {Yeom}(2014)}]{Sasaki:2014spa}%
  \BibitemOpen
  \bibfield  {author} {\bibinfo {author} {\bibfnamefont {M.}~\bibnamefont
  {Sasaki}}\ and\ \bibinfo {author} {\bibfnamefont {D.-h.}\ \bibnamefont
  {Yeom}},\ }\href {\doibase 10.1007/JHEP12(2014)155} {\bibfield  {journal}
  {\bibinfo  {journal} {JHEP}\ }\textbf {\bibinfo {volume} {12}},\ \bibinfo
  {pages} {155} (\bibinfo {year} {2014})},\ \Eprint
  {http://arxiv.org/abs/1404.1565} {arXiv:1404.1565 [hep-th]} \BibitemShut
  {NoStop}%
\bibitem [{\citenamefont {Page}(1976)}]{Page:1976df}%
  \BibitemOpen
  \bibfield  {author} {\bibinfo {author} {\bibfnamefont {D.~N.}\ \bibnamefont
  {Page}},\ }\href {\doibase 10.1103/PhysRevD.13.198} {\bibfield  {journal}
  {\bibinfo  {journal} {Phys.Rev.}\ }\textbf {\bibinfo {volume} {D13}},\
  \bibinfo {pages} {198} (\bibinfo {year} {1976})}\BibitemShut {NoStop}%
\bibitem [{\citenamefont {Myers}\ and\ \citenamefont
  {Perry}(1986)}]{Myers:1986un}%
  \BibitemOpen
  \bibfield  {author} {\bibinfo {author} {\bibfnamefont {R.~C.}\ \bibnamefont
  {Myers}}\ and\ \bibinfo {author} {\bibfnamefont {M.}~\bibnamefont {Perry}},\
  }\href {\doibase 10.1016/0003-4916(86)90186-7} {\bibfield  {journal}
  {\bibinfo  {journal} {Annals Phys.}\ }\textbf {\bibinfo {volume} {172}},\
  \bibinfo {pages} {304} (\bibinfo {year} {1986})}\BibitemShut {NoStop}%
\bibitem [{\citenamefont {Kanti}\ and\ \citenamefont
  {Winstanley}(2015)}]{Kanti:2014vsa}%
  \BibitemOpen
  \bibfield  {author} {\bibinfo {author} {\bibfnamefont {P.}~\bibnamefont
  {Kanti}}\ and\ \bibinfo {author} {\bibfnamefont {E.}~\bibnamefont
  {Winstanley}},\ }\href {\doibase 10.1007/978-3-319-10852-0_8} {\bibfield
  {journal} {\bibinfo  {journal} {Fundam.Theor.Phys.}\ }\textbf {\bibinfo
  {volume} {178}},\ \bibinfo {pages} {229} (\bibinfo {year} {2015})},\ \Eprint
  {http://arxiv.org/abs/1402.3952} {arXiv:1402.3952 [hep-th]} \BibitemShut
  {NoStop}%
\bibitem [{\citenamefont {Hut}\ and\ \citenamefont {Rees}(1983)}]{Hut1983}%
  \BibitemOpen
  \bibfield  {author} {\bibinfo {author} {\bibfnamefont {P.}~\bibnamefont
  {Hut}}\ and\ \bibinfo {author} {\bibfnamefont {M.}~\bibnamefont {Rees}},\
  }\href@noop {} {\bibfield  {journal} {\bibinfo  {journal} {Nature}\ }\textbf
  {\bibinfo {volume} {302}},\ \bibinfo {pages} {508} (\bibinfo {year}
  {1983})}\BibitemShut {NoStop}%
\end{thebibliography}

%

\end{document}